\documentclass[10pt,pre,aps,twocolumn,showpacs,longbibliography]{revtex4-1}
\usepackage{epsfig,amsmath,amssymb,graphicx,color,calc,epstopdf}
\usepackage{color}
\usepackage{inputenc,bm}
\usepackage{hyperref}
\usepackage[mathscr]{euscript}

\newcommand{\revision}[1]{{#1}}
\usepackage{amsmath,amssymb}

\usepackage{changepage}

\usepackage{textcomp,marvosym}

\usepackage{nameref,hyperref}

\usepackage[right]{lineno}

\usepackage{microtype}
\DisableLigatures[f]{encoding = *, family = * }

\usepackage[table]{xcolor}

\usepackage{array}

\usepackage{todonotes}
\usepackage{url}

\usepackage[aboveskip=1pt,labelfont=bf,labelsep=period,justification=raggedright,singlelinecheck=off]{caption}
\usepackage[caption=false]{subfig}

\makeatletter
\renewcommand{\@biblabel}[1]{\quad#1.}
\makeatother

\begin{document}
\title{Classification of crystallization outcomes using deep convolutional neural networks}
\author{Andrew E.~Bruno}
\affiliation{Center for Computational Research, University at Buffalo, Buffalo, New York, United States of America.}
\author{Patrick Charbonneau}
\affiliation{Department of Chemistry, Duke University, Durham,
	North Carolina, USA}
\affiliation{Department of Physics, Duke University, Durham,
	North Carolina, USA}
\author{Janet Newman}
\affiliation{Collaborative Crystallisation Centre CSIRO, Parkville, Victoria, Australia.}
\author{Edward H.~Snell}
\affiliation{Hauptman-Woodward Medical Research Institute and SUNY Buffalo, Department of Materials, Design, and Innovation, Buffalo, New York 14203, United States of America.}
\author{David R.~So}
\affiliation{Google Brain, Google Inc., Mountain View, California, United States of America.}
\author{Vincent Vanhoucke}
\affiliation{Google Brain, Google Inc., Mountain View, California, United States of America.}
\email{vanhoucke@google.com}
\author{Christopher J. Watkins}
\affiliation{IM\&T Scientific Computing, CSIRO, Clayton South, Victoria, Australia}
\author{Shawn Williams}
\affiliation{Platform Technology and Sciences, GlaxoSmithKline Inc., Collegeville, Pennsylvania, United States of America.}
\author{Julie Wilson}
\affiliation{Department of Mathematics, University of York, York, United Kingdom.}

\begin{abstract}
The Machine Recognition of Crystallization Outcomes (MARCO) initiative has assembled roughly half a million annotated images of macromolecular crystallization experiments from various sources and setups. Here, state-of-the-art machine learning algorithms are trained and tested on different parts of this data set. We find that more than 94\% of the test images can be correctly labeled, irrespective of their experimental origin. Because crystal recognition is key to high-density screening and the systematic analysis of crystallization experiments, this approach opens the door to both industrial and fundamental research applications.\\
\textbf{Author summary:}
Protein crystal growth experiments are routinely imaged, but the mass of accumulated data is difficult to manage and analyze. Using state-of-the-art machine learning algorithms on a large and diverse set of reference images, we manage to recapitulate the labels of a remarkably large fraction of the set. This automation should enable a number of industrial and fundamental applications.

\end{abstract}

\date{\today}
\maketitle

\section{Introduction}

X-ray crystallography provides the atomic structure of molecules and molecular complexes. These structures in turn provide insight into the molecular driving forces for small molecule binding, protein-protein interactions, supramolecular assembly and other biomolecular processes. The technique is thus foundational to molecular modeling and design. Beyond the obvious importance of structure information for understanding and altering the role of biomolecules, it also has important industrial applications. The pharmaceutical industry, for instance, uses structures to guide chemistry as part of a ``predict first'' strategy \cite{PredictFirst}, employing expert systems to reduce optimization cycle times and more effectively bring medicine to patients.  Yet, despite decades of methodological advances, crystallizing molecular targets of interest remains the bottleneck of the entire crystallography program in structural biology.

Even when crystallization is facile, it is microscopically rare; for macromolecules it is also uncommon \cite{mcpherson:1999,chayen:2004,fusco:2016,Ng:2016}.  Experimental trials typically involve: (i) mixing a purified sample with chemical cocktails designed to promote molecular association, (ii) generating a supersaturated solution of the desired molecule via evaporation or equilibration, and (iii) visually monitoring the outcomes, before (iv) optimizing those conditions and analyzing the resultant crystal with an X-ray beam. One hopes for the formation of a crystal instead of non-specific (amorphous) precipitates or of nothing at all. In order to help run these trials, commercial crystallization screens have been developed; each screen generally contains 96 formulations designed to promote crystal growth.  Whether these screens are equally effective or not \cite{Ng:2016,Fazio2015141} remains debated, but their overall yield is in any case paltry.
Typically fewer than 5\% of crystallization attempts produce useful results (with a success rate as low as 0.2\% in some contexts~\cite{Newman:2012}).

The practical solution to this hurdle has been to increase the convenience and number of crystallization trials.  To offset the expense of reagents and scientist time, labs routinely employ industrial robotic liquid handlers, nanoliter-size drops, and record trial outcomes using automated imaging systems \cite{HTXIm_WWW,NEWMAN201173,ZHANG201718,Ng:2016,Thielmann201263}. Hoping to compensate for the rarity of crystallization, commercially available systems readily probe a large area of chemical space with minimal sample volume with a throughput of $\sim1000$ individual experiments per hour. 

While liquid handling is readily automated, crystal recognition is not.  Imaging systems may have made viewing results more comfortable than bending over a microscope, but crystallographers still manually inspect images and/or drops, looking for crystals or, more commonly, conditions that are likely to produce good crystals when optimized.  This human cost makes crystal recognition a key experimental bottleneck within the larger challenge of crystallizing biomolecules \cite{Newman:2012}. A typical experiment for a given sample includes four 96-well screens at two temperatures, i.e., 768 conditions (and can have up to twice that~\cite{Snell:2008b}). Assuming that it takes 2 seconds to manually scan a droplet (and noting that the scans have to be repeated, as crystallization is time dependent), simply looking at a single set of 96 trials over the lifetime of an experiment can take the better part of an hour~\footnote{This estimate is based on personnal communication with five experienced crystallographers at GlaxoSmithKline: 2 seconds/observation $\times$ 8 observations $\times$ 96 wells. Note that current technology can  automatically store and image plates at about 3 min/plate.}. For the sake of illustration, the U.S.~Structural Science group at GlaxoSmithKline performs $\sim1200$ 96-well experiments per year.  If the targeted observation schedule were rigorously followed, the group would spend a quarter of the year staring at drops, of which the vast majority contains no crystal. Recording outcomes and analyzing the results of the 96 trials would further increase the time burden. Current operations are already straining existing resources, and the approach simply does not scale for proposed higher-density screening \cite{ZHANG201718}. 
 
Crystal growth is also sufficiently uncommon that the tolerance for false negatives is almost nil. Yet most crystallographers are misguided in thinking that they themselves would never miss identifying a crystal given an image containing an crystal, or indeed miss a crystal in a droplet viewed directly under a microscope~\cite{Wilson:2006}. In fact, not only do crystallographers miss crystals due to lack of attention through boredom, they often disagree on the class an image should be assigned to. An overall agreement rate of $\sim 70\%$ was found when the classes assigned to 1200 images by 16 crystallographers were compared~\cite{Wilson:2006}. (When considering only crystalline outcomes, agreement rose to $\sim 93\%$.) Consistency in visual scoring was also considered by Snell et al.~when compiling a $\sim 150,000$ image dataset~\cite{Snell:2008}. They found that viewers give different scores to the same image on different occasions during the study, with the average agreement rate for scores on a control set at the beginning and middle of the study being 77\%, rising to 84\% for the agreement in scores between the middle and end of the study. Crystallographers also tend to be optimistically biased when scoring their own experiments~\cite{Hargreaves:2018}. A better use of expert time and attention would be to focus on scientific inquiry.  

An algorithm that could analyze images of drops, distinguish crystals from trivial outcomes, and reduce the effort spent cataloging failure, would present clear value both to the discipline and to industry.  Ideally, such an algorithm would act like an experienced crystallographer in:
\begin{itemize}
\item recognizing macromolecular crystals appropriate for diffraction experiments;
\item recognizing outcomes that, while requiring optimization, would lead to crystals for diffraction experiments;
\item recognizing non-macromolecular crystals;
\item ignoring technical failures;
\item identifying non-crystalline outcomes that require follow up;
\item being agnostic as to the imaging platform used;
\item being indefatigable and unbiased;
\item occurring in a time frame that does not impede the process;
\item learning from experience.
\end{itemize}
Such an algorithm would further reduce the variance in the assessments, irrespective of its accuracy. A high-variance, manual process is not conducive to automating the quality control of the system end-to-end, including the imaging equipment. Enhanced reproducibility enables traceability of the outcomes, and paves the way for putting in place measurable, continuous improvement processes across the entire imaging chain.

Automated crystallization image classifications that attempt to meet the above criteria have been previously attempted. The research laboratories that first automated crystallization inspection quickly realized that image analysis would be a huge problem, and concomitantly developed algorithms to interpret them \cite{Spraggon:2002,Cumbaa:2005,Kawabata:2008,Buchala:2008}.
None of these programs was ever widely adopted. This may have been due in part to their dependence on a particular imaging system, and to the relatively limited use of imaging systems at the time. Many of the early image analysis programs further required very time consuming collation of features and significant preprocessing, e.g., drop segmentation to locate the experimental droplet within the image. To the best of our knowledge, there was also no widespread effort to make a widely available image analysis package in the same way that that the diffraction oriented programs have been organized, e.g., the CCP4 package~\cite{Winn:2011}.

Can a better algorithm be constructed and trained?  In order to help answer this question, the Machine Recognition of Crystallization Outcomes (MARCO) initiative was set up~\cite{Marco:2017}. MARCO assembled a set of roughly half a million classified images of crystallization trials through an international collaboration with five separate institutions.  Here, we present a machine-learning based approach to categorize these images. Remarkably, the algorithm we employ manages to obtain an accuracy exceeding 94\%, which is even above what was once thought possible for human categorization. This suggests that a deployment of this technology in a variety of laboratory settings is now conceivable. The rest of this paper is as follows.  Section~\ref{sec:materials} describes the dataset and the scoring scheme, Sec.~\ref{sec:MLmodel} describes the machine-learning model and training procedure, Secs.~\ref{sec:results} and~\ref{sec:discussion} describe and discuss the results, respectively, and Sec.~\ref{sec:conclusion} briefly concludes.

\bigskip

\section{Material and Methods}
\label{sec:materials}

\subsection*{Image Data}
\begin{table*}
\centering
\caption{
{\bf Breakdown of data sources and imaging technology} per institution contributing to MARCO.}
\label{table:7}
\begin{tabular}{|c|c|c|}
\hline
Institution&Technical Setup&\# of Images\\
\hline
Bristol-Myers Squibb&Formulatrix Rock Imager (FRI)&8719\\
CSIRO&Sitting drop, FRI, Rigaku Minstrel~\cite{Vallotton:2010,Rosa:2018}&15933\\
HWMRI&Under oil, Home system~\cite{Snell:2008}&79632\\
GlaxoSmithKline&Sitting drop, FRI&83126\\
Merck&Sitting drop, FRI&305804\\
\hline
\end{tabular}
\end{table*}
The MARCO data set used in this study contains 493,214 scored images from five institutions (See Table~\ref{table:7}~\cite{Marco:2017}). The images were collected from imagers made from two different manufacturers (Rigaku Automation and Formulatrix), which have different optical systems, as well as by the in-house imaging equipment built at the Hauptman-Woodward Medical Research Institute (HWMRI) High-Throughput Crystallization Center (HTCC). Different versions of the setups were also used -- some Rigaku images are collected with a true color camera, some are collected as greyscale images. The zoom extent varies, with some imagers set up to collect a field-of-view (FOV) of only the experimental droplet, and some set for the FOV to encompass a larger area of the experimental setup. The Rigaku and Formulatrix automation imaged vapor diffusion based experiments while the HTCC system imaged microbatch-under-oil experiments. A random selection of 50,284 test images was held out for validation. Images in the test set were not represented in the training set. The precise data split is available from the MARCO website~\cite{Marco:2017}.
 
\subsection*{Labeling}
Images were scored by one or more crystallographers. As the dataset is composed of archival data, no common scoring system was imposed, nor were exemplar images distributed to the various contributors.  Instead, existing scores were collapsed into four comprehensive and fairly robust categories: clear, precipitate, crystal, and other.  This last category was originally used as a catchall for images not obviously falling into the three major classes, and came to assume a functional significance as the classification process was further investigated.  Examination of the least classifiable five percent of images indeed revealed many instances of process failure, such as dispensing errors or illumination problems. These uninterpretable images were then labelled as ``other'' during the rescoring, which added an element of quality control to the overall process~\cite{Mele:2014}.

\subsection*{Relabeling}
After a first baseline system was trained (see Sec.~\ref{sec:MLmodel}), the 5\% of the images that were most in disagreement with the classifier (independently of whether the image was in the training or the test set), were relabeled by one expert, in order to obtain a systematic eye on the most problematic images. 

Because no rules were established and no exemplars were circulated prior to the initial scoring, individual viewpoints varied on classifying certain outcomes. 
For example, the bottom 5\% contained many instances of phase separation, where the protein forms oil droplets or an oily film that coats the bottom of the crystallization well.  Images were found to be inconsistently scored as ``clear'', ``precipitate'', or ``other'' depending on the amount and visibility of the oil film.  This example highlights the difficulty of scoring experimental outcomes beyond crystal identification. A more serious source of ambiguity arises from process failure.  Many of the problematic images did not capture experimental results at all.  They were out of focus, dark, overexposed, dropless, etc. Whatever labeling convention was initially followed, for the relabeling the ``other'' category was deemed to also diagnose problems with the imaging process.  

A total of 42.6\% of annotations for the images that were revisited disagreed with the original label, suggesting somewhat high (1 to 2\%) label noise in this difficult fraction of the dataset. For a fraction of this data, multiple raters were asked to label the images independently and had an inter-rater disagreement rate of approximately 22\%. The inherent difficulty of assigning a label to a small fraction of the images is therefore consistent with the results of Ref.~\cite{Wilson:2006}. Table~\ref{table:labels} shows the final image counts after relabeling.

\begin{table}[!ht]
\centering
\caption{
{\bf Data distribution.} Final number of images in the dataset for each category after collapsing the labels and relabeling.}
\label{table:labels}
\begin{tabular}{|c|c|c|}
\hline
&\multicolumn{2}{c|}{Number of images}\\
\cline{2-3}
Label&Training&Validation\\
\hline
Crystals&56,672&6632\\
\hline
Precipitate&212,541&23,892\\
\hline
Clear&148,861&16760\\
\hline
Other&24,856&3,000\\
\hline
\end{tabular}
\end{table}

\section{Machine Learning Model}
\label{sec:MLmodel}

\begin{figure}
\center{\includegraphics[width=\columnwidth]{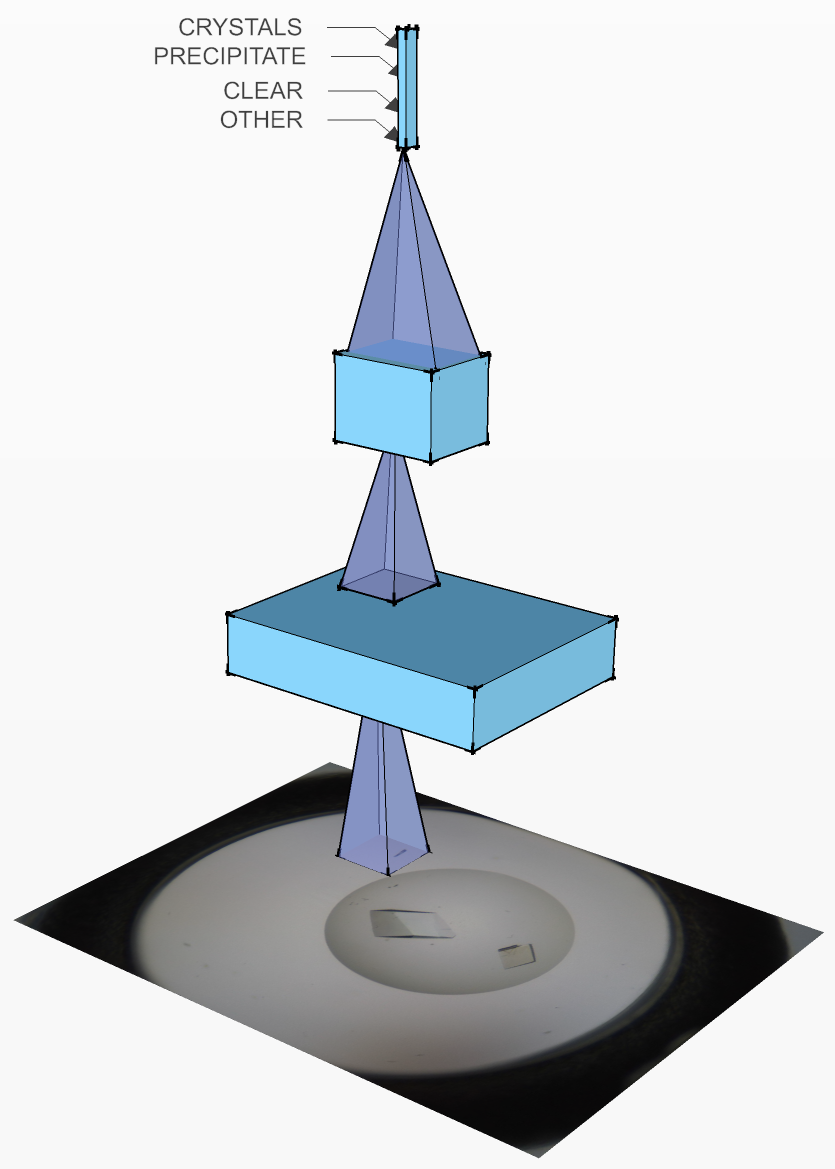}}
\caption{{\bf Conceptual Representation of a Convolutional Neural Network.}
A CNN is a stack of nonlinear filters \revision{(three filter levels are depicted here)} that progressively reduce the spatial extent of the image, while increasing the number of filter outputs that describe the image at every location. On top of this stack sits a multinomial logistic regression classifier, which maps the representation to one probability value per output class (Crystals vs. Precipitate vs. Clear vs. Others). The entire network is jointly optimized through backpropagation~\cite{rumelhart1986learning}, in general by means of a variant of stochastic gradient descent~\cite{bottou2010large}.}
\label{Fig1}
\end{figure}

The goal of the classifier here is to take an image as an input, and output the probability of it belonging to each of four classes (crystals, precipitate, clear, other) (see Fig.~\ref{Fig1}). The classifier used is a deep Convolutional Neural Network (CNN). CNNs, originally proposed in Ref.~\cite{lecun1989backpropagation}, and their modern `deep' variants (see, e.g., Refs.~\cite{lecun2015deep,rawat2017deep} for recent reviews), have proven to consistently provide reliable results on a broad variety of visual recognition tasks, and are particularly amenable to addressing data-rich problems. They have been, for instance, state of the art on the very competitive ILSVRC image recognition challenge~\cite{berg2010large} since 2012.

This approach to visual perception has been making unprecedented inroads in areas such as medical imaging~\cite{litjens2017survey} and computational biology~\cite{angermueller2016deep}, and have also shown to be human-competitive on a variety of specialized visual identification~\cite{krause2017grader,liu2017detecting}. The chosen classifier is thus well suited for the current analysis.

\subsection*{Model Architecture}

The model is a variation on the widely-used Inception-v3 architecture~\cite{szegedy2016rethinking}, which was state of the art on the ILSVRC challenge around 2015. Several more recent alternatives were tried, including Inception-ResNet-v2~\cite{szegedy2017inception}, and automatically generated variants of NASNet~\cite{zoph2017learning}, but none yielded any significant improvements. An extensive hyperparameter search was also conducted using Vizier~\cite{golovin2017google}, also without providing significant improvement over the baseline.

The Inception-v3 architecture is a complex deep CNN architecture described in detail in Ref.~\cite{szegedy2016rethinking} as well as the reference implementation~\cite{slim}. We only describe here the modifications made to tailor the model to the task at hand.

Standard Inception-v3 operates on a 299x299 square image. Because the current problem involves very detailed, thin structures, it is plausible to assume that a larger input image may yield better outcomes. We use instead 599x599 images, and compress them down to 299x299 using an additional convolutional layer at the very bottom of the network, before the layer labeled \texttt{Conv2d\_1a\_3x3} in the reference implementation. The additional convolutional layer has a depth (number of filters) of 16, a $3\times3$ receptive field (it operates on a $3\times3$ square patch convolved over the image) and a stride of 2 (it skips over every other location in the image to reduce the dimensionality of the feature map). This modification improved classification absolute accuracy by approximately 0.3\%. A few other convolutional layers were shrunk compared to the standard Inception-v3 by capping their depth as described in Table~\ref{table:1}, using the conventions from the reference implementation.

\begin{table}[!ht]
\centering
\caption{
{\bf Limits applied to layer depths to reduce the model complexity.} In each named layer of the deep network -- here named after the conventions of the reference implementation -- every convolutional subblock had its number of filters reduced to contain no more than these many outputs.}
\label{table:1}
\begin{tabular}{|c|c|}
\hline
Layer&Max depth\\
\hline
\texttt{Conv2d\_4a\_3x3}&144\\
\texttt{Mixed\_6b}&128\\
\texttt{Mixed\_6c}&144\\
\texttt{Mixed\_6d}&144\\
\texttt{Mixed\_6e}&96\\
\texttt{Mixed\_7a}&96\\
\texttt{Mixed\_7b}&192\\
\texttt{Mixed\_7c}&192\\
\hline
\end{tabular}
\end{table}

While these parameters are exhaustively reported here to ensure reproducibility of the results, their fine tuning is not essential to maximizing the success rate, and was mainly motivated by improving the speed of training. In the end, it was possible to train the model at larger batch size (64 instead of 32) and still fit within the memory of a NVidia K80 GPU (see more details in the training section below). Given the large number of examples available, all dropout~\cite{srivastava2014dropout} regularizers were removed from the model definition at no cost in performance.

\subsection*{Data Preprocessing and Augmentation}

The source data is partitioned randomly into 415990 training images and 47062 test images.

The training data is generated dynamically by taking random 599x599 patches of the input images, and subjecting them to a wide array of photometric distortions, identical to the reference implementation:
\begin{itemize}
\item randomized brightness ($\pm$ 32 out of 255),
\item randomized saturation (from 50\% to 150\%),
\item randomized hue ($\pm$ 0.2 out of 0.5),
\item randomized contrast (from 50\% to 150\%).
\end{itemize}

In addition, images are randomly flipped left to right with 50\% probability, and, in contrast to the usual practice for natural scenes which don't have a vertical symmetry, they are also flipped upside down with 50\% probability. Because images in this dataset have full rotational invariance, one could also consider rotations beyond the mere 90$^{\circ}$, 180$^{\circ}$, 270$^{\circ}$ that these flips provide, but we didn't attempt it here, as we surmise the incremental benefits would likely be minimal for the additional computational cost. This form of aggressive data augmentation greatly improves the robustness of image classifiers, and partly alleviates the need for large quantities of human labels.

For evaluation, no distortion is applied. The test images are center cropped and resized to 599x599.

\subsection*{Training}

\begin{figure}
\center{\includegraphics[width=\columnwidth]{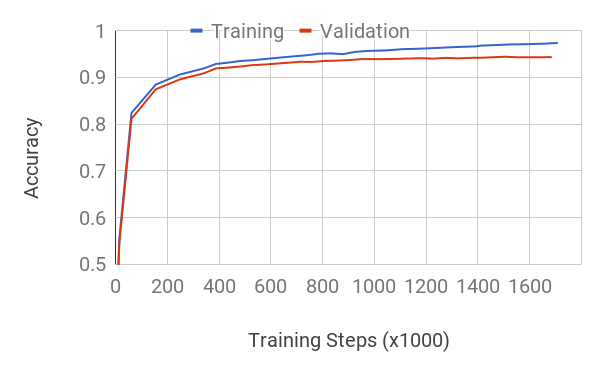}}
\caption{{\bf Classifier Accuracy.}
Accuracy on the training and validation sets as a function of the number of steps of training. Training halts when the performance on the evaluation set no longer increases (`early stopping'). As is typical for this type of stochastic training, performance increases rapidly at first as large training steps are taken, and slows down as the learning rate is annealed and the model fine-tunes its weights.}
\label{Fig2}
\end{figure}

The model is implemented in TensorFlow~\cite{tensorflow2015-whitepaper}, and trained using an asynchronous distributed training setup~\cite{dean2012large} across 50 NVidia K80 GPUs. The optimizer is RmsProp~\cite{tieleman2012lecture}, with a batch size of 64, a learning rate of 0.045, a momentum of 0.9, a decay of 0.9 and an epsilon of 0.1. The learning rate is decayed every two epochs by a factor of 0.94. Training completed after 1.7M steps (Fig.~\ref{Fig2}) in approximately 19 hours, having processed 100M images, which is the equivalent of 260 epochs. The model thus sees every training sample 260 times on average, with a different crop and set of distortions applied each time. The model used at test time is a running average of the training model over a short window to help stabilize the predictions.

\section{Results}

\label{sec:results}

\subsection*{Classification}

The original labeling gave rise to a model with 94.2\% accuracy on the test set. Relabeling improved reported classification accuracy by approximately 0.3\% absolute, with the caveat that the figures are not precisely comparable since some of the test labels changed in between.
The revised model thus achieves 94.5\% accuracy on the test set for the four-way classification task. It overfits modestly to the training set, reaching just above 97\% at the early-stopping mark of 1.7M steps. Table \ref{table:2} summarizes the confusions between classes. \revision{Although the classifier does not perform equally well on images from the various datasets, the standard deviation in performance from one set to another is fairly small, about 5\% (see Table \ref{table:3}), compared to the overall performance of the classifier.}

\begin{table}
\centering
\caption{
{\bf Confusion Matrix.} Fraction of the test data that is assigned to each class based on the posterior probability assigned by the classifier. For instance, 0.8\% of images labeled as Precipitate in the test set were classified as Crystals.}
\label{table:2}
\begin{tabular}{|c|c|c|c|c|}
\hline
True&\multicolumn{4}{c|}{Predictions}\\
\cline{2-5}
Label&Crystals&Precipitate&Clear&Other\\
\hline
Crystals&{\bf 91.0\%}&5.8\%&1.7\%&1.5\%\\
\hline
Precipitate&0.8\%&{\bf 96.1\%}&2.3\%&0.7\%\\
\hline
Clear&0.2\%&1.8\%&{\bf 97.9\%}&0.2\%\\
\hline
Other&4.8\%&19.7\%&5.9\%&{\bf 69.6\%}\\
\hline
\end{tabular}
\end{table}

\begin{table}
\centering
\caption{
\revision{{\bf Standard Deviation} of the predictions across data sources. Note in particular the large variability in the consistency of the label 'Other' across datasets, which leads to comparatively poor selectivity of that less well-defined class.}}
\label{table:3}
\begin{tabular}{|c|c|c|c|c|}
\hline
True&\multicolumn{4}{c|}{Predictions}\\
\cline{2-5}
Label&Crystals&Precipitate&Clear&Other\\
\hline
Crystals&\revision{{\bf     5\%}}& \revision{   4\%}&\revision{     1\%}&  \revision{   1\%}\\
\hline
Precipitate&\revision{     2\%}&\revision{{\bf     4\%}}&\revision{     1\%}&  \revision{   2\%}\\
\hline
Clear& \revision{    1\%}&\revision{     3\%}&\revision{{\bf     5\%}}&  \revision{   1\%}\\
\hline
Other& \revision{    7\%}&\revision{    15\%}& \revision{    6\%}&\revision{{\bf     21\%}}\\
\hline
\end{tabular}
\end{table}

\begin{figure*}
\centerline{
\subfloat[]{\includegraphics[width=\columnwidth]{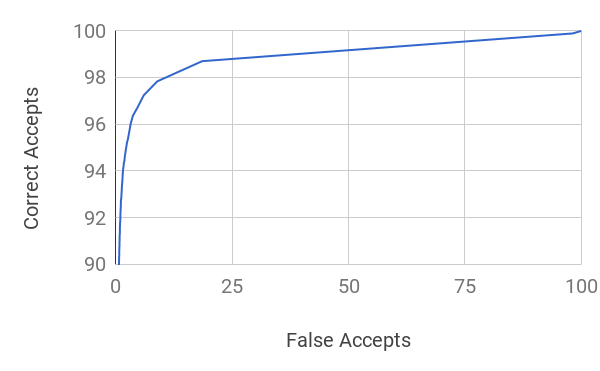}}\quad
\subfloat[]{\includegraphics[width=\columnwidth]{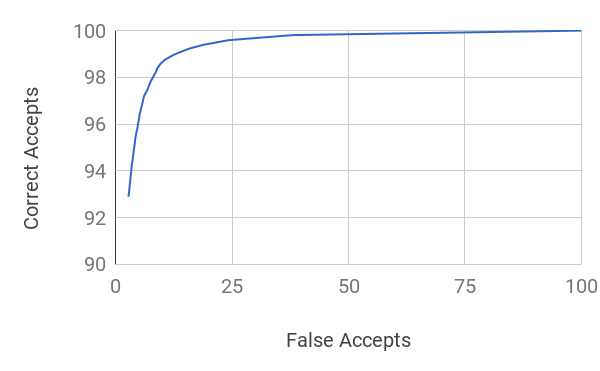}}
}
\caption{{\bf Receiver Operating Characteristic Curves.}
(Q) Percentage of the correctly accepted detection of crystals as a function of the percentage of incorrect detections (AUC: 98.8). 98.7\% of the crystal images can be recalled at the cost of less than 19\% false positives. Alternatively, 94\% of the crystals can be retrieved with less than 1.6\% false positives.
(B) Percentage of the correctly accepted detection of precipitate as a function of the percentage of incorrect detections (AUC: 98.9). 99.6\% of the precipitate images can be recalled at the cost of less than 25\% false positives. Alternatively, 94\% of the precipitates can be retrieved with less than 3.4\% false positives.}
\label{Fig3}
\end{figure*}

The classifier outputs a posterior probability for each class. By varying the acceptance threshold for a proposed classification, one can trade precision of the classification against recall. The receiver operating characteristic (ROC) curves can be seen in Fig.~\ref{Fig3}.
\revision{\subsection*{Validation}
At CSIRO C3 a workflow~\cite{c3pipeline} has been set up which uses a variation of the analysis tool from DeepCrystal~\cite{deepcrystal} to analyze newly collected crystallisation images and to assign either no score, `crystal’ score or `clear’ score. A total of 37,851 images were collected in Q1 2018 and assigned a human score by a C3 user were used as an independent dataset to test the MARCO tool. Within this dataset, 9746 images had been identified as containing crystals. The current, DeepCrystal tool (which assigns only ‘crystal’ or ‘clear’ scores) was found to have an overall accuracy rate of 74\%, while the MARCO tool has 90\%. Although this retrospective analysis doesn’t allow for a direct comparison of the ROC, the precision, recall and accuracy of the two tools all favor the MARCO tool, as shown in table 6. The precision achieved by MARCO on this dataset is also very similar to that seen for the CSIRO images in the training data.}
\begin{table}
\centering
\caption{
\revision{{\bf Validation at C3} Precision, recall and accuracy from an independent set of images collected after the MARCO tool was developed. The 38K images of sitting drop trials were collected between January 1 and March 30, 2018 on two Formulatrix Rock Imager (FRI) instruments.}}
\label{table:6}
\begin{tabular}{|c|c|c|c|}
\hline
DL tool&Precision&Recall&Accuracy\\
\hline
DeepCrystal&0.4928&0.4520&0.7391\\
MARCO&0.7777&0.8663&0.9018\\
\hline
\end{tabular}
\end{table}

\subsection*{Pixel Attribution}

\begin{figure*}
\centerline{
\subfloat[]{\includegraphics[width=0.32\textwidth]{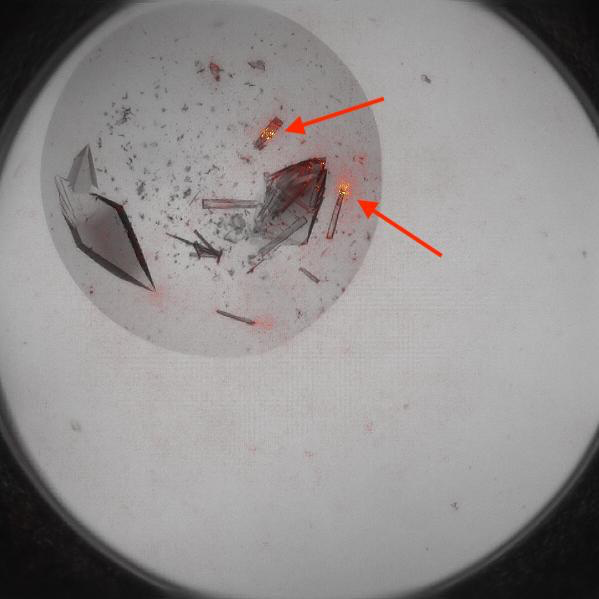}}\quad
\subfloat[]{\includegraphics[width=0.32\textwidth]{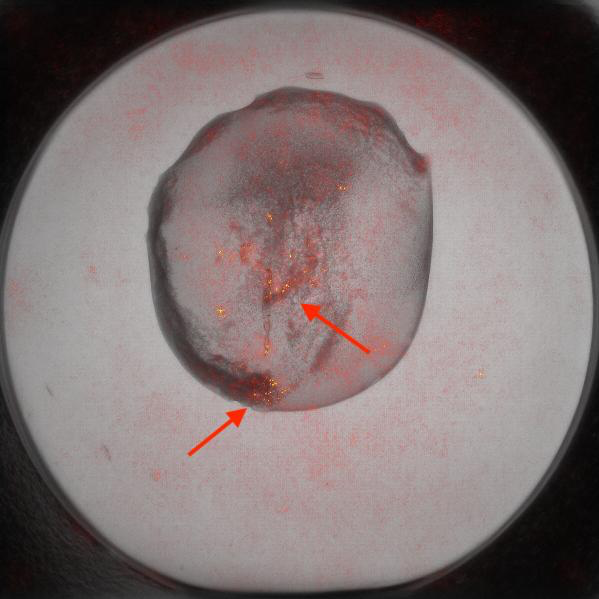}}\quad
\subfloat[]{\includegraphics[width=0.32\textwidth]{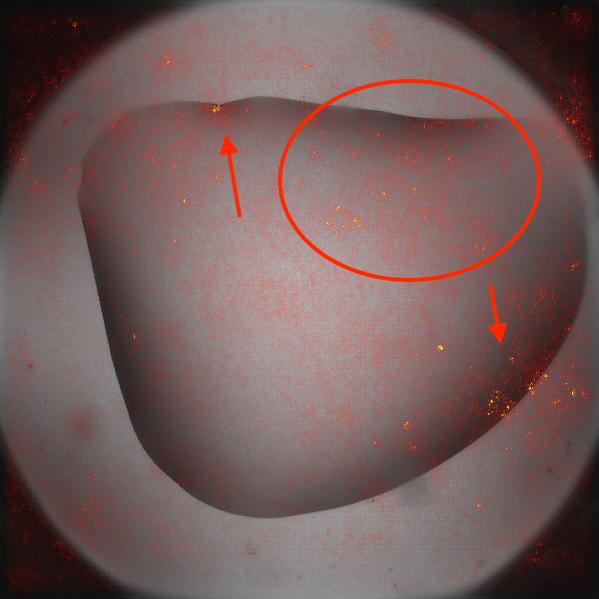}}}
\caption{{\bf Sample heatmaps for various types of images.}
(A) Crystal: the classifier focuses on some of the angular geometric features of individual crystals (arrows). (B) Precipitate: the classifier lands on the precipitate (arrows). (C) Clear: The classifier broadly samples the image, likely because this label is characterized by the absence of structures rather than their presence. Note the slightly more pronounced focus on some darker areas (circle and arrows) that could be confused for crystals or precipitate. \revision{Because the `Others' class is defined negatively by the the image being not identifiable as belonging to the other three classes, heatmaps for images of that class are not particularly informative.}}
\label{Fig4}
\end{figure*}

We visually inspect to what parts of the image the classifier learns to attend by aggregating noisy gradients of the image with respect to its label on a per-pixel basis. The SmoothGrad~\cite{smilkov2017smoothgrad} approach is used to visualize the focus of the classifier. The images in Fig.~\ref{Fig4} are constructed by overlaying a heat map of the classifier's attention over a grayscale version of the input image.

Note that saliency methods are imperfect and do not in general weigh faithfully all the evidence present in an image according to their contributions to the decision, especially when the evidence is highly correlated. Although these visualizations paint a simplified and very partial picture of the classifier's decision mechanisms, they help confirm that it is likely not picking up and overfitting to cues that are irrelevant to the task.

\subsection*{Inference and Availability}

\revision{The model is open-sourced and available online at~\cite{tfmodelsmarco}. It} can be run locally using TensorFlow or TensorFlow Lite, or as a Google Cloud Machine Learning~\cite{cloudml} endpoint. At time of writing, inference on a standard Cloud instance takes approximately 260ms end-to-end per standalone query. However, due to the very efficient parallelism properties of convolutional networks, latency per image can be dramatically cut down for batch requests.

\section{Discussion}
\label{sec:discussion}
Previous attempts at automating the analysis of crystallisation images  have employed various pattern recognition and machine learning techniques, including linear discriminant analysis~\cite{Cumbaa:2003,Saitoh:2005}, decision trees and random forests~\cite{Bern:do5006,Liu20081187,Cumbaa:2010}, and support vector machines~\cite{Pan:2006,Buchala:2008}. Neural networks, including self-organizing maps, have also been used classify these images~\cite{Spraggon:2002,Po20081926}, with the most recent involving deep learning~\cite{Yann:2016}. However, all previous approaches have required a consistent set of images with the same field of view and resolution, in order to identify the crystallization droplet in the well~\cite{Vallotton:2010}, and thereby restrict the analysis. Various statistical, geometric or textural features were then extracted, either directly from the image or from some transformation of the region of interest, to be used as variables in the classification algorithms. 

The results from various studies can be difficult to compare head-to-head because different groups present confusion matrices with the number of classes ranging from 2 to 11, only sometimes aggregating results for crystals/crytalline materials. There is also a tradeoff between the number of false negatives and the number of false positives. Yet most report classification rates for crystals around 80-85\% even in more recent work~\cite{Cumbaa:2010,HTXIm_WWW,Hung:2014}, in which missed crystals are reported with much lower rates. This advance comes at the expense of more false positives. For example, Pan et al.~report just under 3\% false negatives, but almost 38\% false positives~\cite{Pan:2006}.

As the trained algorithms are specific to a set of images, they are also restricted to a particular type of crystallisation experiment. Prior to the curation of the current dataset, the largest set of images (by far) came from the Hauptman-Woodward Medical Research Institute HTCC~\cite{Snell:2008}. This dataset, which contains 147,456 images from 96 different proteins but is limited to experiments with the microbatch-under-oil technique, has been used in a number of studies~\cite{Fusco:2014,Yann:2016}. 
Most notably, Yann et al.~used a deep convolutional neural network that automatically extracted features, and reported a correct classification rates as high as 97\% for crystals and 96\% for non-crystals. Although impressive, these results were however obtained from a curated subset of 85,188 \emph{clean} images, i.e., images with class labels on which several human experts agreed~\cite{Yann:2016}. In order to validate our approach, we retrained our model to perform the same 10-way classification on that subset of the data alone without any tuning of the model's hyperparameters and achieved 94.7\% accuracy, compared to the reported 90.8\%.

In this context, the current results are especially remarkable. A crystallographer can classify images of experiments independently of the systems used to create those images. They can view an experiment with a microscope or look at a computer image and reach similar conclusions. They can look at a vapor diffusion experiment or a microbatch-under-oil setup and, again, asses either with confidence. Here, we show that this can be accomplished equally well, if not better, using deep CNNs. A benchtop researcher can classify many images, especially if they relate to a project that has been years in the making. For high-throughput approaches, however, that task becomes challenging. The strength of computational approaches is that each image is treated like the previous one, with no fatigue. Classification of 10,000 images is as consistent as classification of one. This advance opens the door for complete classification of all results in a high-throughput setting and for data mining of repositories of past image data.

Another remarkable aspect of our results is that they leverage a very generic computer vision architecture originally designed for a different classification problem -- categorization of natural images -- with very distinct characteristics. For instance, one can presume that the global geometric relationships between object parts would play a greater role in identifying a car or a dog in an image, compared to the very local, texture-like features involved in recognizing crystal-like structures. Yet no particular specialization of the model was required to adapt it to the widely differing visual appearances of the samples originating from different imagers. This convergence of approaches toward a unified perception architecture across a wide range of computer vision problems has been a common theme in recent years, further suggesting that the technology is now ready for wide adoption for any human-mediated visual recognition task.

\section{Conclusion}
\label{sec:conclusion}
In this work, we have collated biomolecular crystallization images for nearly half a million of experiments across a large range of conditions, and trained a CNN on the labels of these images. Remarkably, the resulting machine-learning scheme was able to recapitulate the labels of more than 94\% of a test set. Such accuracy has rarely been obtained, and has no equal for an uncurated dataset. The analysis also identified a small subset of problematic images, which upon reconsideration revealed a high level of label discrepancy. This variability inherent to using human labeling highlights one of the main benefits of automatic scoring. Such accuracy also make conceivable high-density screening.

Enhancing the imaging capabilities by including UV or SONICC results, for instance, could certainly enrich the model. But several research avenues could also be pursued without additional laboratory equipment. In particular, it should be possible to leverage side information that is currently not being used.
\begin{itemize}
\item The four-way classification scheme used is a distillation of 38 categories which are present in the source data. While these categories are presumed to be somewhat inconsistent across datasets, they could potentially provide an additional supervision signal.
\item Because one goal of this classifier is to be able to generalize {\it across} datasets, it would be worthwhile to investigate the contribution of techniques that have been designed to specifically reduce the effect of domain shift across data sources on the classification outcomes~\cite{ganin2015unsupervised,bousmalis2016domain}.
\item Each crystallization experiment records a series of images taken over times. Using the timecourse information could enhance the success rate of the classifier~\cite{Mele:2013}.
\end{itemize}

Note in closing that the current study focused on crystallization as an outcome, which is but a small fraction of the protein solubility diagram. Patterns of precipitation, phase separation, and clear drops, also provide information as to whether and where crystallization might occur. The success in identifying crystals, precipitate and clear can be thus also be used to accurately chart the crystallization regimes and to identify pathways for optimization~\cite{Snell:2008c,Fusco:2014,Altan:2016}.  The application of this approach to large libraries of historical data may therefore reveal patterns that guide future crystallization strategies, including novel chemical screens and mutagenesis programs.

\section*{Acknowledgments}
We acknowledge discussions at various stages of this project with I.~Altan, S.~Bowman, R.~Dorich, D.~Fusco, E.~Gualtieri, R.~Judge, A.~Narayanaswamy, J.~Noah-Vanhoucke, P.~Orth, M.~Pokross, X.~Qiu, P.~F.~Riley, V.~Shanmugasundaram, B.~Sherborne and F.~von Delft.
PC acknowledges support from National Science Foundation Grant no.~NSF DMR-1749374.

\end{document}